\documentclass[12pt]{article}
\usepackage[pctex32]{graphics}
\newcommand{\be}{\begin{equation}}
\newcommand{\ee}{\end{equation}}
\newcommand{\ba}{\begin{eqnarray}}
\newcommand{\ea}{\end{eqnarray}}

\def\<{\langle}
\def\>{\rangle}

\begin{document}
\begin{center}
{\huge An application of interpolating scaling functions to wave packet
propagation}

\vskip 1cm
{\large Andrei G. BORISOV ${}^{a,}$
\footnote{
E-mail address: {\sf borisov@lcam.u-psud.fr} }
 \ \ and \ \ Sergei V. SHABANOV ${}^{b,}$
\footnote{E-mail address: {\sf shabanov@phys.ufl.edu}}
}

\vskip 1cm
{\it {${}^a$\ Laboratoire des Collisions Atomiques et Mol\'eculaires,
UMR CNRS-Universit\'{e} Paris-Sud 8625, B\^{a}t. 351,
Universit\'{e} Paris-Sud, 91405 Orsay CEDEX, France 

${}^b$\ Department of Mathematics, University of Florida, 
Gainesville, FL 23611, USA
}}
\end{center}

\vskip 2cm
\begin{abstract}
Wave packet propagation in the basis of  
interpolating scaling functions (ISF) is studied. The ISF are well known
in the multiresolution analysis based on spline biorthogonal wavelets.
The ISF form a cardinal basis set
corresponding to an equidistantly spaced grid. They have 
compact support of the size determined by the order of the underlying
interpolating polynomial that is used to generate ISF. 
In this basis
the potential energy matrix is diagonal and the
kinetic energy matrix is sparse and, in the 1D case, has a band-diagonal
structure. An important feature of the basis is that 
matrix elements of a Hamiltonian are exactly computed by means of
simple algebraic transformations efficiently
implemented numerically. Therefore the number of grid points and the
order of the underlying interpolating polynomial can easily be varied
allowing one to approach the accuracy of pseudospectral methods in a regular
manner, similar to high order finite difference methods. The results of
numerical simulation of a H+H$_{2}$ collinear collision show that the ISF
provide one with an accurate and efficient representation for use in the wave
packet propagation method.
\end{abstract}

\newpage

\section{Introduction}

A direct on-the-grid solution of the time-dependent Schr\"{o}dinger equation
(TDSE) has become a common tool in quantum chemistry. In a 
dynamical context, it
provides with quantitative predictions on the efficiency of 
different pathways of chemical reactions and deepens our understanding of their
details \cite{1,2,3}. Approaches based on the TDSE in connection with the
filter diagonalization technique \cite{4} are also used in the static
context in order to compute states of complex molecules \cite{5,6}.
To simulate the time evolution of a wave packet, 
repeated computations 
of the  action of the system  Hamiltonian $H$, or its exponential
$\exp (-i\Delta tH)$, on the wave function $\Psi $ are required. Here $\Delta
t$ is a time step. Therefore a lot of effort has been
devoted to developing accurate and efficient methods 
to reduce computational costs of computing
$H\Psi $. The numerical techniques used can be classified
into several categories: (i) finite differences (FD), (ii) finite elements
(FE), (iii) pseudospectral global grid representation approaches such as the
discrete variable representation (DVR) and Fourier grid Hamiltonian method
(FGH) \cite{7,8,9,10,11}. Finite differences and finite elements lead to a
sparse Hamiltonian matrix but exhibit 
a slow algebraic convergence with the number of
grid points. The pseudospectral approaches result in dense Hamiltonian
matrices,
which, a priori, increases the number of operations needed to compute 
$H\Psi $. At the same time, the exponential convergence with the number of grid
points counterbalances the aforementioned drawback. This is why 
pseudospectral methods are so widely used in time-dependent molecular
dynamics \cite{1,2,3,12}, as well as in stationary S-matrix \cite{13,14,15} and
eigenvalue \cite{6} calculations. In particular, the FGH method based on the
fast Fourier transform (FFT) algorithm is very advantageous since for a
mesh of $N$ points the action of the kinetic energy operator is computed
by $N\log_2 N$ elementary multiplications 
and is easily implemented numerically \cite{16}.

When discussing the slow algebraic convergence of finite differences and
finite elements, one usually refers to the convergence with the number of grid
points while the order of finite differences or the order of the polynomial for
finite elements is kept fixed. In fact, an exponential (spectral)
convergence can be achieved if not only the number of grid points, but
also the order of the underlying polynomial is increased \cite{7,17}. 
Exploiting this property, as well as the sparsity of 
the Hamiltonian matrix and the possibility
to distribute mesh points only along the reaction path, thus reducing the
size of calculation, makes the high order FD and FE a competitive
alternative to pseudospectral methods. Much work has recently been devoted
to developing these techniques \cite{18,19,20,21,22,23,24,25}.

In this paper we present a treatment of reactive scattering based on
the wave function representation in the basis of interpolating scaling
functions corresponding to interpolating (spline) wavelets \cite{28}. 
Basis functions are generated from a single
function, called the scaling function, by appropriate scalings and
shifts of its argument. The scaling function is a solution 
of a functional equation which is found iteratively by using
interpolating polynomials of a specific order. By construction,
the basis functions have compact
support with the width determined by the order of the interpolating
polynomial and the resolution level (the mesh step). The basis is
biorthogonal and cardinal, which leads to a simple representation of the wave
function. The resulting Hamiltonian matrix is sparse. 
The potential energy matrix appears to be diagonal, while,
in the one dimensional case, the kinetic energy matrix is band
diagonal with the band width determined  by the size of support of the
scaling function. Matrix elements of the
kinetic energy operator can be evaluated exactly via simple algebraic
operations so that the order of the underlying interpolating polynomial and
the step of the mesh can easily be varied allowing one to achieve a fast
convergence. Thus our algorithm offers flexibility similar to that
achieved in the case of finite differences of an arbitrary order with the
Fornberg algorithm \cite{7,26}. In fact, the proposed approach can be seen
as an alternative to high-order finite difference techniques.
Applications of the method are illustrated with the example 
of a collinear H+H$_{2}$ collision.

\section{Theory}

\subsection{Biorthogonal spline bases}

Here we summarize some necessary facts about interpolating
biorthogonal bases of scaling functions in the space of square integrable
functions and describe the algorithm to construct such bases. 
Biorthonormal bases of scaling functions are used in the  
multiresolution analysis associated with interpolating wavelets. 
A more detailed description of such bases can be found in the
mathematical literature, e.g.,  \cite{27,28,29}. 

In general, a biorthogonal basis consists of two sets of
elements $\phi _{a}(x)$ and $\tilde{\phi}_{a}(x)$ where the index $a$ labels
the basis elements. Any function $\Psi (x)$ can be decomposed into a linear
combination of the basis functions $\phi _{a}(x)$, 
\begin{equation}
\label{1}
\Psi (x)=\sum_{a}s_{a}\phi _{a}(x)\ ,  
\end{equation}
where the decomposition coefficients
are determined by the dual basis
functions 
\begin{equation}
\label{2}
s_{a}=\int dx\,\tilde{\phi}_{a}(x)\Psi (x)\ .  
\end{equation}
The basis is (bi)orthogonal in the sense that 
\begin{equation}
\label{3}
\int dx\,\tilde{\phi}_{a}(x)\phi _{b}(x)=\delta _{ab}\ .  
\end{equation}
Consider a special class of biorthogonal bases that provide
a multiresolution analysis. A biorthogonal basis with a multiresolution
analysis is generated by a scaling function $\phi (x)$ and its dual $\tilde{%
\phi}(x)$ which satisfy accordingly the equations (the scaling relations)
\ba
\label{4}
\phi (x)&=&2\sum_{k}h_{k}\phi (2x-k)\ , \\ 
\label{5}
\tilde{\phi}(x)&=&2\sum_{k}\tilde{h}_{k}\tilde{\phi}(2x-k)\ ,
\ea
where the real coefficients $h_{k}$ ($\tilde{h}_{k}$) are called a (dual)
filter. The scaling function and its dual are required to be orthogonal in
the sense that 
\begin{equation}
\label{or}
\int dx\,\tilde{\phi}(x)\phi (x-j)=\delta _{0j} 
\end{equation}
for all integers $j$. The orthogonality relation imposes a condition on the
filters which is readily deduced from (\ref{or}) by substituting
the scaling relations (\ref{4}) and (\ref{5}) are rescaling the
integraion variable.

Consider two sets of functions, labeled by two integers $n$ and $j$, 
\begin{equation}
\label{7}
\phi _{n,j}(x)=2^{n/2}\phi (2^{n}x-j)\ ,\ \ \ \ \ \tilde{\phi}
_{n,j}(x)=2^{n/2}\tilde{\phi}(2^{n}x-j)\ .  
\end{equation}
Subspaces $V_{n}$ of the space of square integrable functions spanned by $%
\phi _{n,j}$ with a fixed value of $n$ form a ladder structure 
\begin{equation}
\label{8}
\cdots \subseteq V_{n}\subseteq V_{n+1}\subseteq \cdots \ .  
\end{equation}
It is straightforward to convince oneself that $\phi _{n,j}$ form a
biorthogonal basis in $V_{n}$: 
\begin{equation}
\label{9}
\int dx\,\tilde{\phi}_{n,j}(x)
\phi _{n,j^{\prime }}(x)=\delta _{jj^{\prime }}\ . 
\end{equation}
Any function $\Psi $ can be projected on a subspace $V_{n}$, 
\begin{equation}
\label{proj}
P_{n}:\ \ \ \Psi(x) \rightarrow \Psi _{n}(x)
=\sum_{j}s_{n,j}\phi _{n,j}(x)\ ,\ \ \ \
s_{n,j}=\int dx\,\tilde{\phi}_{n,j}(x)\Psi(x) \ .  
\end{equation}
Taking successively larger values of $n$ allows one to reproduce a
successively finer structure of $\Psi $. Thus, the index $n$ specifies a
resolution level. If the filters are finite, then the scaling function $\phi 
$ has finite support and so do the basis functions $\phi _{n,j}$. The index $%
j$ is then naturally associated with the 
position of support of $\phi _{n,j}$.

A special class of biorthogonal bases is obtained when filters are
finite and of a special form 
\ba
\label{11}
h_{k}&=&\frac{1}{2}\phi (k/2)\ ,\ \ \ \ \ k=0,\pm 1,\pm 2,...,\pm m\ ,\\
\label{12}
\tilde{h}_{k}&=&\delta_{k0}\ ,\ \ \ \ \ \ \
\tilde{\phi}(x)=\delta (x)\ . 
\ea
The expansion coefficients are simply values of the function at
dyadic lattice sites 
\begin{equation}
\label{13}
s_{n,j}=2^{-n/2}\Psi (2^{-n}j)\ .  
\end{equation}
Larger values of the resolution level $n$ correspond to finer grids.
In what follows $m$ is chosen to be odd for convenience.

To find the filter $h_{k}$ and an explicit form of the scaling function $\phi
(x)$, Eq. (\ref{4}) is solved iteratively for $\phi (x)$.
First, one observes that the equation is satisfied at the 
integer valued argument by $\phi (l)=\delta _{0l}$
with $h_{0}=\frac{1}{2}$. Thus, the scaling function vanishes at integral $x$ 
except for $x=0$ where $\phi (0)=1$ as is required by the
orthogonality condition (\ref{or}) with our choice of the filters
(\ref{11}) and (\ref{12}). 
To compute $\phi
(x)$ at half-integer values of the argument, $\phi (j+1/2)$ ($j$ is a fixed
integer), one uses the polynomial (spline) interpolation with 
polynomials $P_{m}^{(j)}$ of order $m$ passing through $M=(m+1)/2$ integer
points neighboring to $x=j+1/2$ to the left and $M$ integer points to the
right: $x=j-M+1,j-M+2,...,j+M$. 
Having found such a polynomial, we set $\phi
(j+1/2)=P_{m}^{(j)}(j+1/2)$. Note that for different points $x=j+1/2$,
polynomials $P_{m}^{(j)}$ are different. To find $\phi (j\pm 1/4)$, the same
procedure is applied, but now the values of $\phi $ at $M$ half-integer
points neighboring to $x=j\pm 1/4$ to the left and $M$ half-integer points
to the right are used to construct the corresponding interpolating
polynomial. For example, for $x=j+1/4$ the sequence will be: $%
x=j-(M-1)/2,j-(M-2)/2,...,j+M/2$. In other words, any $x$ can be squeezed
into successively smaller intervals of length $1/2,\ 1/4,\ ...,\ 1/2^{N}$, $%
N\rightarrow \infty $. The limiting procedure allows one to compute $\phi
(x) $ at any $x$, in principle.

An important property of $\phi (x)$ is that it has compact support,
an interval of width $D=2m$, that is,
$\phi (x)=0$ for all $|x|\geq m$ as can be inferred from the
construction procedure described above.

In Fig.~1 we show an  example of the scaling function $\phi (x)\equiv \phi
_{0,0}(x)$ where $\phi _{0,0}(x)$ is the basis function centered at position 
$x=0$ and corresponding to the zero resolution level (the mesh mesh equals
$1$). The length of the filter is $m=15$ so that support of the
scaling function is the interval
$-15\leq x\leq 15$.

\subsection{Hamiltonian matrix}

Consider first a simple $1D$\ example. Let $H$ be a Hamiltonian
of a system in the coordinate representation. A solution of the Schr\"odinger
equation $i\partial/\partial t\ \Psi
(x,t)=H\Psi (x,t)$ for a given initial wave packet $\Psi(x,t=0)$ is
approximated by its
projection into a finite dimensional subspace
spanned by $\phi _{n,j}$ where $j$ enumerates basis functions whose support
lies in a box, $x\in \lbrack 0,L]$, that is, $\Psi(x,t) = \sum_j
s_{n,j}(t) \phi_{n,t}(x)$.
Each basis function
$\phi _{n,j}$ has support of the length $D_{n}=2m/2^{n}$. Therefore 
the number of the basis functions $%
N $ is given by: $N=2^{n}L/2m$. Our choice of the basis implies also
zero boundary conditions for the wave function.
The initial wave packet
$\Psi(x,t=0)$ is projected into the corresponding subspace of $V_n$
according to (\ref{proj}) to determine $s_{n,j}(0)$. The Hamiltonian 
is projected by the rule $H\rightarrow P_nHP_n$ and becomes a finite
matrix with elements
\be
\label{14}
H_{kj}^{(n)}=\int dx\,\tilde{\phi}_{n,k}(x)H\phi _{n,j}(x)\ . 
\ee
The Hamiltonian matrix acts in a vector space of the expansion 
coefficient $s_{n,j}(t)$. Solving the time dependent Schr\"odinger 
equation thus implies computing $s_n(t) = \exp(-itH^{(n)}) s_n(0)$,
where $s_n$ is regarded as a vector with components $s_{n,j}$ and
$H^{(n)}$ as a matrix with elements given in (\ref{14}). 
 
A typical Hamiltonian is a sum of kinetic and potential energies. For matrix
elements of the potential energy $V$ we have 
\begin{equation}
\label{15}
V_{kj}^{(n)}=
\int dx\,\tilde{\phi}_{n,j}(x)V(x)\phi _{n,j}(x)=V(2^{-n}j)\delta
_{kj}\ .  
\end{equation}
This is a great advantage of the basis under consideration: The potential
energy matrix is diagonal. To compute the kinetic energy matrix we apply a
general procedure to compute derivative
operators $d^{l}/dx\,^{l}$ in a basis of compactly supported (spline) wavelets 
\cite{30}. By rescaling the integration variable, and using 
Eqs. (\ref{7}) and (\ref{12}) one can
find 
\begin{eqnarray}
D_{k,j}^{(n,l)} &=&\int dx\,\tilde{\phi}_{n,k}(x)(d/dx\,)^{l}
\phi _{n,j}(x) 
=2^{nl}\int dx\,\tilde{\phi}(x)(d/dx\,)^{l}\phi
(x-k+j)\nonumber \\ \label{16}
&=&2^{nl}(d/dx\,)^{l}\phi (x-k+j)\mid _{x=0}\equiv
2^{nl}D_{j-k}^{(l)}\ .
\end{eqnarray}
Using the scaling relation for the scaling function (\ref{4}) we infer 
\begin{eqnarray}
\label{17}
D_{i}^{(l)} &=&2^{l+1}\sum_{k}D_{2i-k}^{(l)}h_{k}\equiv
\sum_{j}A_{ij}^{(l)}D_{j}^{(l)}\ ,   \\ \label{18}
A_{ij}^{(l)} &=&2^{l+1}h_{2i-j}=2^{l}\phi (i-j/2)  \ .
\end{eqnarray}
Thus, $D^{(l)}$ is an eigenvector of the matrix $A^{(l)}$ corresponding to
the eigenvalue $1$.

Finally, we need a normalization of the vector $D^{(l)}$. Note that, since
the value of $\phi $ at any $x$ is given by a polynomial of order $m$, the
monomial $x^{l}$, $l\leq m$, should be a linear combination of $\phi
_{0,j}(x)=\phi (x-j)$, that is, $x^{l}=\sum_{j}s_{0,j}\phi _{0,j}(x)$ with $%
s_{0,j}=j^{l}$. Differentiating this relation $l$ times, multiplying by $%
\tilde{\phi}(x)$ and integrating over $x$ we obtain the normalization
relation 
\begin{equation}
\label{19}
l!=\sum_{j}j^{l}D_{j}^{(l)}\ .  
\end{equation}
The matrix $D^{(n,l)}$ satisfies the symmetry relation $%
D_{kj}^{(n,l)}=(-1)^{l}D_{jk}^{(n,l)}$, which follows from (\ref{16}) after
changing the integration variable $x\rightarrow -x$ and making use of $\phi
(x)=\phi (-x)$ (the same for the dual). In particular, $D^{(n,2)}$ is a
symmetric matrix. Since support of $\phi (x)$ lies within $|x|\leq m$, the $%
D^{(n,l)}$ matrix is band diagonal: $D_{k,j}^{(n,l)}=0$, if $\left|
k-j\right| >m$ as follows from Eq. (\ref{16}). Therefore the action of the
Hamiltonian of the system $H^{(n)}=-\frac{1}{2}D^{(n,2)}+V^{(n)}$ 
on the wave function $s_n$
requires $N\times 2m$ elementary multiplications. The above approach allows easy and fast
evaluation of the Hamiltonian matrix for any desirable resolution level $n$
(any number of the basis functions) and any length of the filter $m$ (any
interpolating polynomial order).

A multidimensional generalization is obtained by taking the direct product
of $\phi _{n,j}$ for every independent variable $x_{i}$. The resolution
level $n_{i}$ may be chosen independently for every variable $x_{i}$. For
example, in the two dimensional case, the basis consists of functions $\phi
_{n_{1},j_{1}}(x_{1})\phi _{n_{2},j_{2}}(x_{2})$. The spline order $m$ may
also be chosen independently for every coordinate. In other words, the
scaling functions for each $x_{i}$ may be different.

\section{Numerical results}

\subsection{Harmonic oscillator example}

In Fig. 2 we show the results of a test calculation of  $30$ first
eigenvalues of a 1D harmonic oscillator,  $H=-
\frac{1}{2}\frac{d^{2}}{dx^{2}}+\frac{x^{2}}{2}$. We use a mesh
in the interval $-L\leq
x\leq L$ with $L=10$ $a_{0}$ where $a_0$ is the Bohr radius. 
The Hamiltonian matrix in the basis of
interpolating scaling functions  $\phi _{n,j}(x)$ has been calculated
according to the procedure described in the previous section and diagonalized
yielding the eigenstate energies. The results of the present approach are
compared with those obtained by the Fourier Grid Hamiltonian (FGH)
approach. In the FGH method, the convergence is reached with 80 points of the
grid and the error of the eigenvalue calculation is basically
determined by the
precision of the diagonalization procedure. In the ISF basis and low order $%
m,$ convergence with the number of basis functions $N$ ($N=2^{n}\times
2L $, where $n$ is the resolution level) is slow. At the same time the
convergence can dramatically be improved by increasing the order $m$ of
the interpolating polynomial, i.e. by increasing the band width of the
band-diagonal kinetic energy matrix. This observation is in line with the
results reported by several authors for finite differences and finite
elements where it has been shown that the pseudospectral convergence can be
reached by increasing the order of finite differences or the order of an
underlying polynomial in finite elements \cite{1,17,18,19,25}. Thanks
to the biorthogonality of the basis in our case,
increasing the order $m$ and the resolution level $n$ is a quite
simple procedure given by Eqs. (\ref{15})--(\ref{19}) which is easy 
to implement numerically. 

In the case of the harmonic oscillator the FGH method
outperforms the ISF method. The latter, as well as finite
difference methods, might compete with pseudospectral approaches for 
calculations involving complex reaction paths. Indeed, the FGH method, for
example, generally uses hypercubic grid domains. In the present approach
the calculation volume can sufficiently be reduced by the choice of the basis
functions $\phi _{n,j}$ whose support lies in the vicinity of the reaction
path, which is again a simple procedure because of the biorthogonality
of the basis. We now turn to one such example.

\subsection{H+H$_{2}(v=0)$ collinear collision}

Here we present a wave packet propagation treatment
of an H+H$_{2}$ collision in collinear geometry with the
energy of the collisional Hydrogen atom between $0.2\, eV$ and $1.1\, eV$. 
The system is described by
Jacobi coordinates $r_{1,2}$ with $r_{1}$ being the distance between the two
Hydrogen atoms in the molecule and $r_{2}$ being the distance of the
collisional Hydrogen atom from the molecular center of mass. We use the
Lanczos method \cite{31,32} to solve the time dependent Schr\"{o}dinger
equation (atomic units are used)
\begin{equation}
\label{20}
i\, \frac{\partial}{\partial t}\ \Psi (r_{1},r_{2};t)
=\left( -\frac{1}{2m_H}\left[
2\,\frac{\partial ^{2}}{\partial r_{1}^{2}}+\frac{3}{2}\frac{\partial ^{2}}{
\partial r_{2}^{2}}\right] +V(r_{1},r_{2})\right) \Psi (r_{1},r_{2};t)
\ ,
\end{equation}
where $m_H$ is the mass of the Hydrogen atom. The interaction potential $%
V(r_{1},r_{2})$ is taken from \cite{33}. A typical size of the mesh is $%
32\times 32$ $a_{0}$. An absorbing potential \cite{34,35} is introduced at the
grid boundaries for large $r_{1}$ and $r_{2}$ to avoid
interference of the simulated wave packet with its reflection 
from the grid boundary. The initial state corresponds to
the H$_{2}$ molecule in the $v=0$ vibrational state and an impinging Gaussian
wave packet in the reaction channel $(r_{2})$. 

In the FGH method, convergent
results are obtained with $192$ points in the $r_{1}$ coordinate and $256$
points in the $r_{2}$ coordinate. Two types of calculation have been
performed by the ISF method. The first one corresponds
to a rectangular grid where the basis functions are chosen as $\Phi
_{n,j_{1}j_{2}}=\phi _{n,j_{1}}(r_{1})\phi _{n,j_{2}}(r_{2})$. The
resolution level $n=4$ has been used, which corresponds 
to the mesh step $1/2^{4}=0.0625$ $a_{0}$ so that in total 
there are $360\times 512$ basis functions.
The order of the underlying interpolating polynomial (size of the filter)
has been set to $m=21$. 
In the second simulation the ISF basis has been
chosen in such a way that the basis functions have support in 
the potential energy valley inside the $%
4.6$ $eV$ potential energy level curve as shown in Fig. 3. In this case
only $48600$ basis functions are needed. As a result, the
computational time has been reduced by
$3.8$ times, which brings it to the level comparable to the FGH treatment.
Results are presented in Fig. 4. We find that with our choice of the
basis both simulations based on the interpolating scaling functions 
yield the total reaction probability $R$ which nicely agrees
with the FGH method. The absolute difference $\left| \frac{R_{FGH}-R_{ISF}}{%
R_{FGH}}\right| $ is better than $0.1\%$.

\section{Conclusions}

We have shown that the wave function representation in  bases of 
interpolating scaling functions can efficiently be used in  wave packet
propagation studies of molecular dynamics. The ISF originate from a
multiresolution analysis associated with interpolating (spline) 
wavelets \cite{27,28,29}. The
ISF have a finite support whose width is determined 
by the order of the underlying
interpolating polynomial and the resolution level (the mesh step). The
basis is cardinal, which leads to a simple representation of the wave function.
The resulting Hamiltonian matrix is sparse. In particular, 
the potential energy matrix is diagonal, while
for the kinetic energy matrix there is an efficient and simple
algebraic procedure for its exact computation 
so that the order of the underlying
interpolating polynomial and the mesh step can easily be varied
to achieve fast convergence. Thus, our algorithm provides one with
flexibility similar to that achieved in the case of finite differences of
an arbitrary order by, e.g., the  Fornberg algorithm \cite{7,26}. 
In the 1D case the kinetic energy matrix has a band diagonal structure with the
width of the band of off-diagonal elements given by the size of
support of the scaling function. In our example
of a collinear H+H$_{2}$ collision, a considerable reduction of 
computational costs has been reached when the basis functions are
 chosen in such a way that their support is localized  only in the
vicinity  of the reaction path. The very possibility of the latter
procedure
and the simplicity of its numerical implementation, thanks to the
biorthogonality of the basis, 
is a general feature of the method proposed which can be used 
in simulations of systems with more complex reaction paths 
to reduce computational costs. 

\vskip 1cm
{\large\bf Acknowledgemets}

S.V.S. akcnowledges the support of LCAM (University of Paris-Sud) and is grateful for
the warm hospitality during his stay in Orsay where most of this
work has been done. We thank J.R. Klauder for reading the paper and
useful remarks.

\vskip 1cm

{\bf Figure captions}

Fig. 1.\qquad Scaling function $\phi (x)$ for the filter length $m=15$%
.\bigskip

Fig. 2\qquad A 1D harmonic oscillator problem. A relative error in the
eigenvalue calculation as a function of the 
oscillator quantum number. 
Lines: The FGH
method with different numbers $N$ of mesh points; the gray
line is for $
N=40$, the dashed line for $N=80$, and the black line for $N=160$. 
Symbols and lines with
symbols: The ISF method. 
Circles show the results obtained with $n=2$ resolution level
corresponding to the $N=80$ scaling functions basis ($N=80$ points of the
mesh). Triangles show the results obtained with $n=3$ resolution level
corresponding to the $N=160$ scaling functions basis ($N=160$ points of the
mesh). Gray solid symbols: Results obtained with the filter length $m=7$
(the 
interpolating polynomial order); open symbols: Results 
obtained with the filter
length $m=15$; black solid symbols: Results obtained with the filter length
$m=21$. The line with black solid symbols: 
results obtained with the filter length $m=41$.\bigskip

Fig. 3\qquad A schematic representation of the arrangement of 
of the mesh (positions of the basis functions) 
in the wave packet propagation treatment 
of a collinear H-H$_{2}$ collision.\bigskip

Fig. 4.\qquad Reaction probability calculated by different methods for the
H+H$_{2}(v=0)$ collinear collision as a function of energy. Solid line:
Fourier Grid Hamiltonian method; Open circles: Results obtained by ISF
method in the rectangular mesh; Triangles: Results obtained by the 
ISF method in
the mesh arranged along the reaction coordinate (path) as depicted in Fig. 3.

\end{document}